\documentclass{article}
\usepackage{emulateapj}
\usepackage{apjfonts}
\usepackage{epsfig}

\newcommand{\fb}{$f_{bar}$}
\newcommand{\ba}{$(b/a)_{\rm bar}$}

\begin{document}


\title{Explorations in Hubble Space: A Quantitative Tuning Fork}

\author{Roberto G. Abraham}
\affil{Department of Astronomy, University of Toronto, 60 St.
George Street, Toronto, ON, M5S 3H8, Canada.
abraham@astro.utoronto.ca}

\medskip

\author{Michael R. Merrifield}
\affil{School of Physics \& Astronomy, University of Nottingham,
Nottingham NG7 2RD, UK. michael.merrifield@nottingham.ac.uk.}

\begin{abstract}
In order to establish an objective framework for studying galaxy
morphology, we have developed a quantitative two-parameter description
of galactic structure that maps closely on to Hubble's original tuning
fork.  Any galaxy can be placed in this ``Hubble space,'' where the
$x$-coordinate measures position along the early-to-late sequence, while
the $y$-coordinate measures in a quantitative way the degree to which
the galaxy is barred. The parameters defining Hubble space are
sufficiently robust to allow the formation of Hubble's tuning fork to
be mapped out to high redshifts.  In the present paper, we describe a
preliminary investigation of the distribution of local galaxies in
Hubble space, based on the CCD imaging atlas of Frei et al.\
(1996). We find that barred, weakly-barred, and unbarred galaxies are
remarkably well-separated on this diagnostic diagram.  The spiral
sequence is clearly bimodal and indeed approximates a tuning fork:
strongly-barred and unbarred spirals do not simply constitute the
extrema of a smooth unimodal distribution of bar strength, but rather
populate two parallel sequences.  Strongly barred galaxies lie on a
remarkably tight sequence, strongly suggesting the presence of an
underlying unifying physical process.  Rather surprisingly, weakly
barred systems do not seem to correspond to objects bridging the
parameter space between unbarred and strongly barred galaxies, but
instead form an extension of the regular spiral sequence. This
relation lends support to models in which the bulges of late-type
spirals originate from secular processes driven by bars.
\end{abstract}


\section{INTRODUCTION} \label{sec:introduction}

In any branch of science, classification systems work best when
applied to subsets of objects that are unambiguously distinct. In such
cases, a meaningful classification system can lead to fundamental
insights.  Two obvious examples are Mendeleev's organization of
chemical data into the periodic table of the elements, a discovery
which anticipated electron orbital structure by decades, and Linnean
binary nomenclature, which has formed the basis of the scientific
naming scheme applied to organisms for over two centuries, and which
lies at the foundation of most studies of evolution and natural
selection.

On the other hand, the division of a smoothly varying population into
subjectively defined classes has always proved a ``sure formula for
endless bickering among specialists, for no two will ever agree''
(Gould 1998).  On this basis, it seems reasonable to ask whether a truly
fundamental classification scheme for galaxies will ever be devised,
because, unlike systems whose components are either perfectly discrete
or of roughly uniform strength, galaxy properties span a broad
continuum.  In addition, the complexity of galaxies' morphologies
means that any classification is likely to be based on a rather
complex mixture of properties rather than a single defining feature.
For such systems it may be more meaningful to think in terms of
distributions in a multi-dimensional parameter space rather than in
terms of discrete classes (Abraham et al.\ 1994; Bershady, Jangren, \&
Conselice 2000; van den Bergh et al. 2000).

Despite these caveats, Hubble's (1926) simple scheme for
classifying galaxy morphology has proved remarkably versatile and
robust.  Hubble classified galaxies into ellipticals and spirals,
defining a sequence from ``early'' to ``late'' types\footnote{It
is commonly held that Hubble used this nomenclature because he
believed that the galaxies formed a temporal sequence.  In
fact, he explicitly warns against interpreting his choice of
adjective in this way (Hubble 1926, p. 326).} by the size of the
central bulge, and the smoothness and pitch angle of the spiral
arms, using the letters `a' to `c' to designate the location
along the sequence.  He then further divided the spiral galaxies
into two parallel sequences of barred (`SB') and unbarred (`S')
galaxies, depending on whether the central region contained a
strong non-axisymmetric distortion.  Drawing the full sequence
from elliptical galaxies to late-type spiral galaxies, with the
spiral sequence bifurcating into parallel barred and unbarred
sequences, one obtains the classical ``tuning fork'' scheme for
classifying galaxy morphology.  The catch-all category of
``irregular'' allows one to deal with galaxies that do not fit
within this simple scheme, but in the nearby Universe the vast
majority of bright galaxies can be fairly reliably placed
somewhere on the tuning fork.

In the past decade, a number of factors have arisen that drive us
to look beyond such a qualitative scheme.  Imaging surveys
encompassing millions of galaxies, such as the Sloan Digital Sky
Survey (SDSS), are now sufficiently large that visual
classification is impractical. Furthermore, direct studies of
galaxy evolution are now possible by imaging galaxies at high
redshift using  the {\em Hubble Space Telescope} (HST).  However,
with the notable exception of data from the {\em Hubble Deep
Field} (Williams et al. 1996), HST images of distant galaxies are
not of high enough quality for it to be possible to undertake
traditional morphological classifications directly onto Hubble's
tuning fork, since the finer details of spiral structure cannot
be robustly studied at the signal-to-noise levels typical of
these data.

In order get around this difficulty, the morphologies of distant
galaxies have been probed using several different techniques,
including:
\begin{enumerate}
\item Visual classifications using coarse bins that neglect the visibility
of spiral structure (Griffiths et al. 1994; Glazebrook et al.
1995; Driver et al. 1995; van den Bergh et al. 1996).
\item Classifications based on surface-brightness profile fits to analytical models
(Schade et al 1995; Odewahn et al 1997; Phillips et al. 1997;
Marleau \& Simard 1998; Corbin et al. 2000).
\item More general automated schemes based on various combinations of image
concentration and surface brightness (Abraham et al.\ 1994),
image concentration and asymmetry (Abraham et al.\ 1996;
Brinchmann et al.\ 1998; Volonteri, Saracco, \& Chincarini 2000),
and image concentration, asymmetry,
surface brightness, and spectral class (Bershady, Jangren,
\& Conselice 2000).
\end{enumerate}
\noindent Analysis using such techniques places the
epoch at which the general framework of the early-to-late sequence
is established to be at a redshift of  $z \sim 1$ (Brinchmann et al.
1998; Driver et al. 1998).

Evidence for even more recent evolution in Hubble's tuning fork has
come from studies of the fraction of barred galaxies as a function of
redshift. Unlike other aspects of spiral structure, bars are fairly
high surface brightness features and are visible to quite high
redshifts. van den Bergh et al.\ (1996) noted that the data from the
Hubble Deep Field North gave the visual impression that there are few
barred galaxies at high redshift. A quantitative analysis of both
Hubble Deep Fields by Abraham et al.\ (1999) confirmed this visual
impression, and showed that barred galaxies are rare at redshifts
beyond $z\sim 0.5$. Recent data from the Caltech Faint Galaxy Redshift
Survey (CFGRS) has lent further support to this picture (van den Bergh
et al.\ 2000).

\begin{figure*}[htb]
  \centering \epsfig{figure=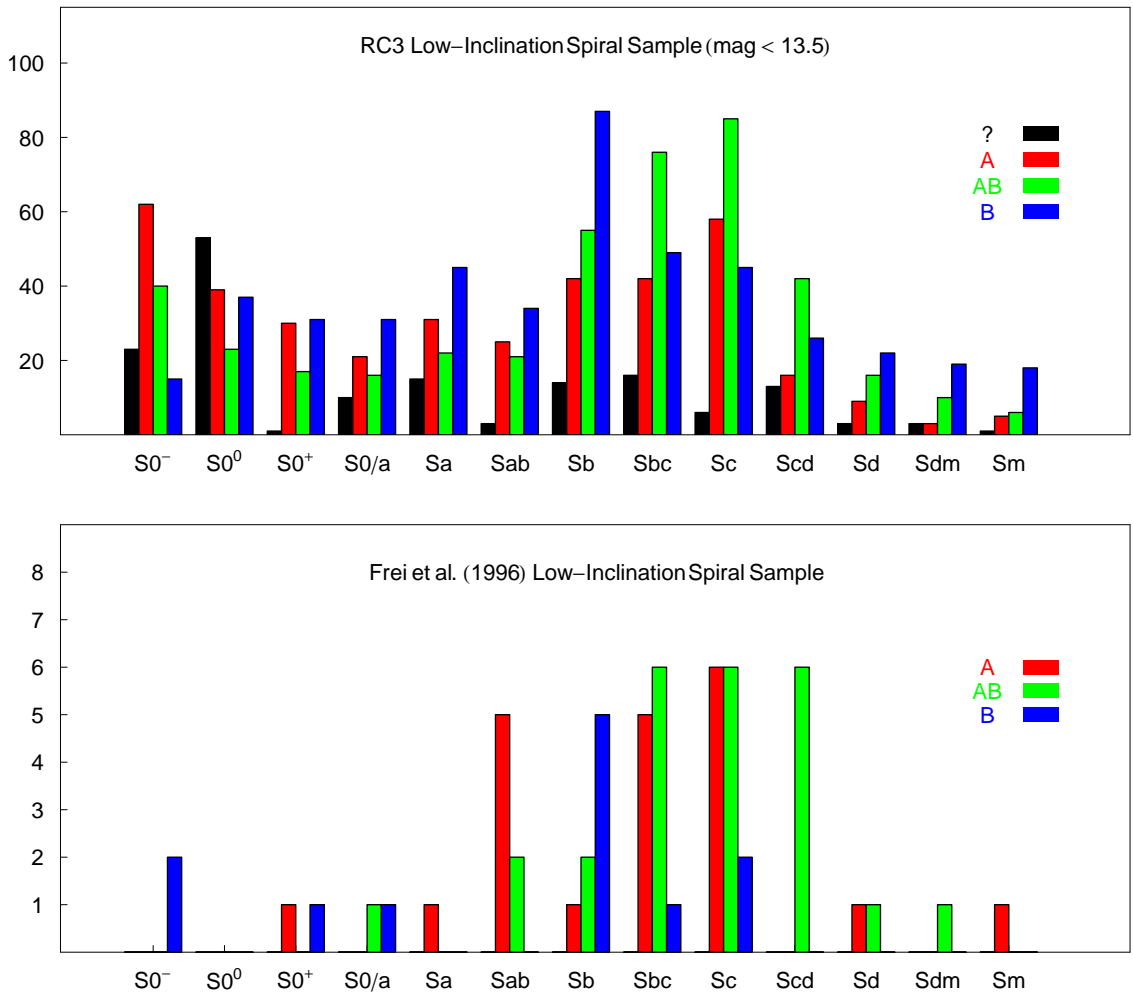,width=6.5in}
  \caption{Distributions of unbarred [red], weakly barred
  [green], and strongly barred [blue] galaxies in the Third
  Reference Catalog [upper panel] and in the Frei et al.\ 1996
  atlas [lower panel]. Numbers of galaxies are shown as a
  function of Hubble stage.  All distributions correspond to
  systems with inclinations less than 60 degrees, as inferred
  from their axial ratios. The RC3 sample was also cut at a
  magnitude limit of $B=13.5$ mag.  Fainter than this limit
  unclassified systems (black histograms) begin to dominate the
  trends.} \label{fig:rc3frei}
\end{figure*}

A major stumbling block in characterizing the evaporation of the
Hubble sequence as a function of redshift has been the absence of
a quantitative framework for describing the classical tuning
fork.  The primary aim of this paper is to set such a framework.
Any galaxy can be placed in this ``Hubble space'', where the
$x$-coordinate measures position along the early-to-late
sequence, while the $y$-coordinate measures in a quantitative way
the degree to which the galaxy is barred.  Since this system is
fully parametric, it is not keyed in any way to local archetypes,
and the system is robust enough to be useful over a broad range
of redshifts.

\begin{figure*}[bht]
  \centering
  \caption{{\bf [See 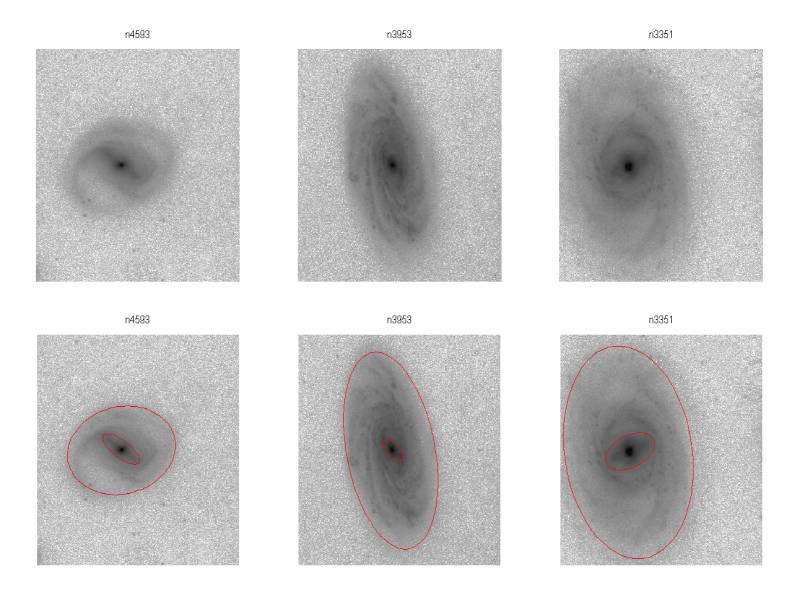 file]}
  [Top] Representative SB galaxies from the sample
  of Frei et al.\ (1996).  [Bottom] The same galaxies shown  with
  superposed inner and outer ellipses defined using the procedure
  described in the text. }
  \label{fig:examples}
\end{figure*}

\begin{figure*}[bht]
  \centering
  \epsfig{figure=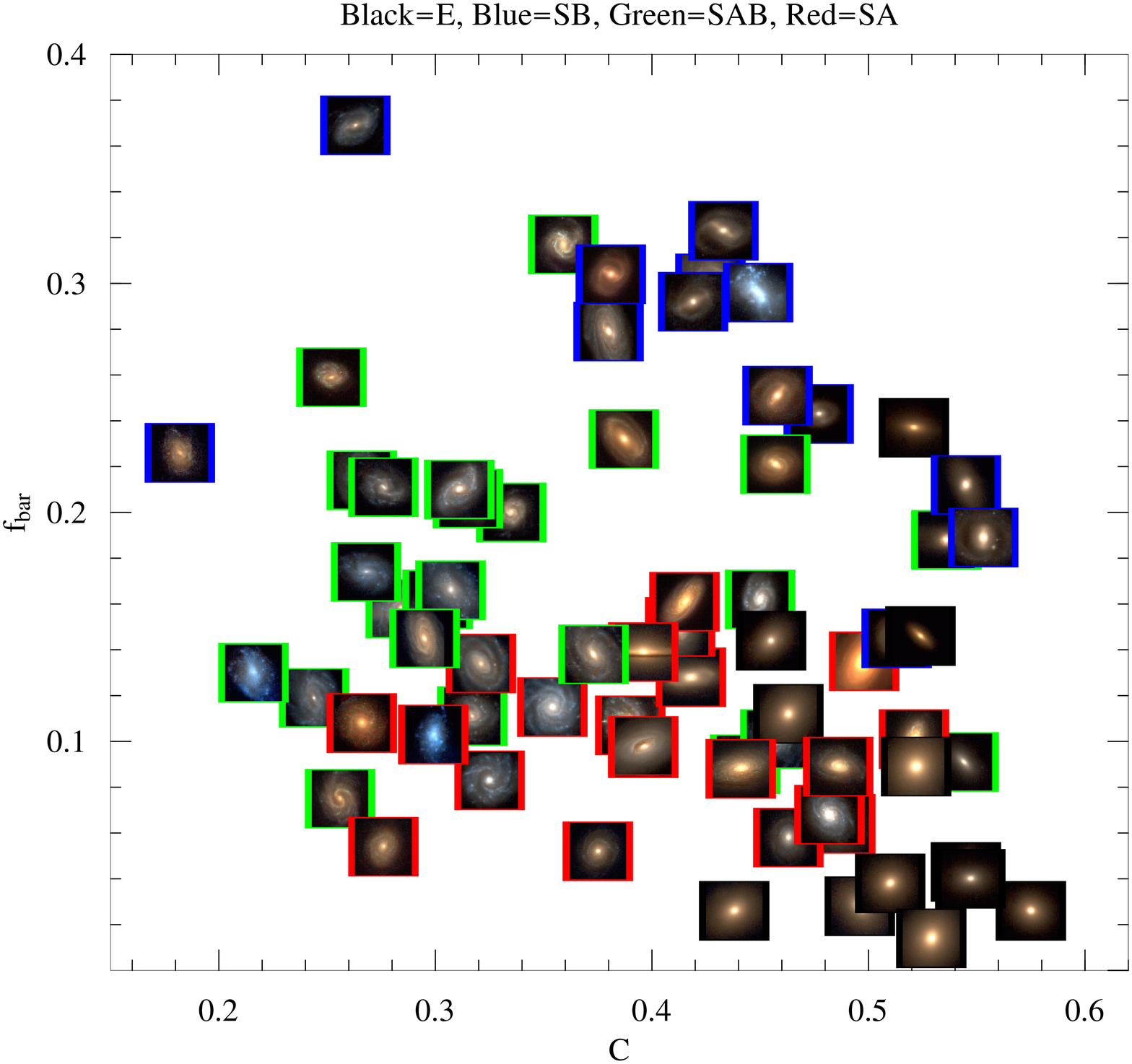,width=6.5in}
\caption{ The distribution of the Frei et al.\ (1996) sample in
Hubble space. ``Postage stamp'' images are shown at the position
of each object in the central concentration versus bar strength
diagram.  Colored boxes subdivide the galaxy population into SA
[red], SAB [green], and SB [blue] spirals.  Elliptical galaxies
are shown in black boxes.} \label{fig:stamps}
\end{figure*}

The second aim of this paper is to apply this quantitative analysis to
the morphological classification of nearby galaxies.  Specifically, we
seek to determine whether a tuning fork-shaped distribution genuinely
defines the local ``morphological zero-point'', or whether Hubble's
classification scheme is only a convenient idealization. The
bifurcation implied by the tuning fork paradigm itself implies a
genuine distinction between barred and unbarred galaxies.  It is
therefore important to determine whether a bimodal distribution of
barred and unbarred galaxies exists in Hubble space, or whether these
objects are better described as corresponding to the tails of a
unimodal distribution, with the ``average'' galaxy lying somewhere in
the middle.  In addition to providing a baseline against which the
properties of distant galaxies can be compared, this analysis may also
offer insights into the physical interpretation of Hubble's original
classification scheme.

The remainder of this paper is arranged as follows.  In
Section~\ref{sec:method}, we present the two quantities adopted to
parameterize Hubble space. In Section~\ref{sec:frei} we make a
preliminary exploration of Hubble space using the data from the
Frei et al.\ (1996) atlas. The limitations of this sample for
conducting this investigation are also discussed.  The
implications of this analysis are discussed in
Section~\ref{sec:discussion}, and our conclusions are summarized
in Section~\ref{sec:conclusions}.

\section{DEFINITION OF HUBBLE SPACE} \label{sec:method}

\subsection{Measuring Hubble Stage}

As described above, the classical Hubble type of a galaxy is
based on the visual impression of several aspects of the system's
properties. The compound nature of this classification scheme is
not conducive to summarizing a galaxy's morphology with a single
quantitative parameter. However, the close correlation between the
parameters defining the Hubble type means that one can obtain a
remarkably good measure of position on the early-to-late sequence
[the ``Hubble stage'' as defined in the {\em Third Reference
Catalog of Bright Galaxies}, hereafter RC3 (de~Vaucouleurs et al.\
(1991)] by focusing on a single parameter, namely the bulge-to-disk
ratio of the galaxy.  We therefore measure the location of
galaxies along the early-to-late axis of Hubble space using the
central concentration parameter, $C$, defined in Abraham et al.\
(1994) and closely related to the parameter defined by Doi,
Fukugita, \& Okamura (1993).  This quantity tracks bulge-to-disk
ratio very closely, and has been shown to provide a good
quantitative substitute for more orthodox visual classifications
(Abraham et al.\ 1996).

\subsection{Measuring Bar Strength}

A first attempt at robustly measuring the orthogonal ordinate in
Hubble space, the bar strength, was made by Abraham et al.\
(1999). Their parameter, \ba, measured the intrinsic axis ratio
of any central bar under the assumption that the distribution of
light corresponds to a thin disk that is intrinsically
axisymmetric at large radii.  This model is clearly an
idealization, but still provides a robust objective measure of
bar strength irrespective of the true distribution of light in
the galaxy.  As noted by Abraham et al.\ (1999), the assumption
of a two-dimensional disk clearly becomes unreasonable if a
galaxy is viewed close to edge-on, when the three-dimensional
shape of the central bulge becomes a prominent feature.  However,
it is intrinsically almost impossible to determine whether a
galaxy that lies close to edge-on is barred on the basis of
photometry alone.  We therefore follow Abraham et al.\ (1999), and
only attempt to determine the value of \ba\ for galaxies whose
axial ratios imply an inclination of $i < 60\,{\rm degrees}$.

Abraham et al.\ (1999) calculated \ba\ using the second order moments
obtained by slicing a galaxy image at two intensity levels. The
moments at these slices were used to define ``inner'' and ``outer''
ellipses whose axis ratios and orientations were used to calculate
\ba.  A deficiency of this technique is the rather {\em ad hoc} choice
of intensity levels used to define these ellipses (10\% and 85\% of
the image maximum after the top 2\% of the image pixels had been
clipped), since it is conceivable that barred systems could exist with
isophotal signatures outside the range explored by these cuts.  To
remedy this deficiency, we here adopt a ``multi-thresholding''
approach to define the optimum intensity cuts, using the following
procedure.  A single outer ellipse is defined using second-order
moments obtained by a 3$\sigma$ cut above the sky noise, $\sigma$.  A
succession of inner ellipses is then defined by slicing the galaxy
image at a range of intensity levels spaced apart by 1$\sigma$.  At
each cut, only those pixels contiguous with the center of the galaxy
(defined as the position of the pixel with maximum flux) are retained.
Some portions of a galaxy will be poorly described by elliptical
isophotes, either intrinsically or because of poor sampling near the
center of the object.  Therefore, ellipses encompassing fewer than
80\% of the pixels contained within the original isophote are
discarded, as are ellipses with semi-major axis lengths smaller than
5\% of the radius of the outer isophote.  For every remaining ellipse,
\ba\ is calculated using equation~(2) of Abraham et al.\ (1999).  The
final value of \ba\ adopted for the galaxy is the minimum value of
\ba\ over all slices, which therefore defines the maximum of any
bar-like distortion in the galaxy.

As a further refinement, we then calculate a quantity that is more
closely related to most morphologists' subjective notion of bar
strength.  Specifically, we define
\begin{equation}
f_{bar} = {2 \over \pi} \left\{\arctan\left[{(b/a)_{bar}}^{-1/2}\right]
                            - \arctan\left[{(b/a)_{bar}}^{1/2}\right]\right\}.
\end{equation}
This parameterization maps the bar strength into a closed interval
from zero (unbarred) to unity (infinitely strong bar).  For an
idealized picture of a galaxy containing a central elliptical bar of
uniform surface brightness, this quantity is the minimum fraction of
the bar's stars that one would have to rearrange in order to transform
the structure into an axisymmetric distribution.  As discussed above,
we are not arguing that such an idealized model represents a real
galaxy in any sense; rather, this explanation is intended to give a
physical insight into the interpretation of the parameter, which
remains valid for real systems.

\begin{figure*}[bht]
  \centering
  \epsfig{figure=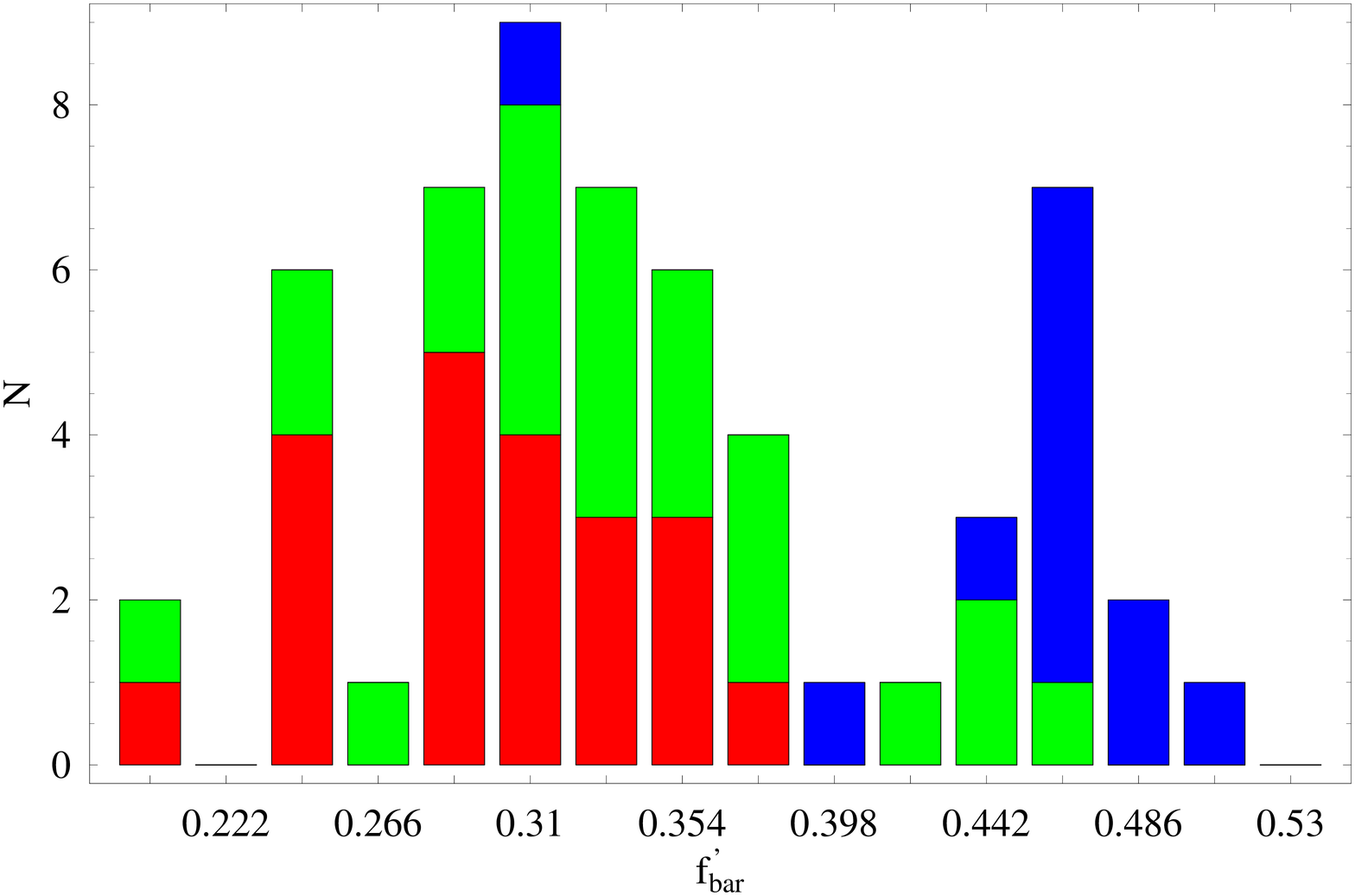,width=6.5in}
  \caption{
  Histogram showing the distribution of spiral galaxies perpendicular to
  the sequences in Fig.~\ref{fig:stamps}.  The color coding is the same
  as in Fig.~\ref{fig:stamps}. The histogram was calculated by
  rotating the distribution clockwise 60 degrees about the point (0.4,0.2)
  and then projecting onto the x-axis.}
\label{fig:hist}
\end{figure*}

\section{THE DISTRIBUTION OF GALAXIES IN HUBBLE SPACE} \label{sec:frei}

Having defined the physical measure of Hubble space, we are now in a
position to see how it is populated by galaxies.  In this preliminary
investigation, we will analyze the sample of nearby galaxies from the
Frei et al.\ (1996) CCD imaging atlas.  Of the 113 galaxies in the
atlas, 91 are spirals, 56 of which satisfy the inclination cut of $i <
60\,{\rm degrees}$. The CCD images in the Frei et al.\ sample were
obtained using two telescopes (the Palomar 1.5m and the Lowell 1.1m)
using different filters.  Since the local morphological distribution
has been defined using blue-sensitive photographic plates, in the
present analysis we adopt the bluest images available for each object
in the atlas ($B_J$-band images for data obtained at Lowell, and
Gunn-$g$ images for observations from Palomar).

It is important to emphasize that the Frei et al.\ (1996) atlas
was not designed to represent a flux-limited sample of objects:
its original purpose was to provide a useful ``training set'' of
galaxies for automated classification in the SDSS.  It is clear
that large homogeneous digital surveys such as the SDSS will
ultimately provide the best way forward toward the overall goal
of understanding the distribution of galaxy morphologies in the
Universe.  However, we will now show that the Frei et al.\ (1996)
sample can already be used to draw several fairly general
conclusions regarding the distribution of galaxies in Hubble
space.

Figure~\ref{fig:rc3frei} compares the distribution of morphological
classifications for all low-inclination $B<13.5$~mag galaxies in the
RC3, along with classifications for the corresponding low-inclination
Frei et al.\ (1996) sub-sample. Individual histograms for the various
bar classes from the RC3 are shown.  Here, we adopt the terminology of
the RC3, which denotes strongly barred spirals as class SB,
weakly/tentatively barred systems as class SAB, and unbarred spirals
as class SA, with lower-case Roman letters suffixed to denote Hubble
stage.  Clearly, the Frei et al.\ (1996) sample is strongly biased
against early-type disk systems (S0 galaxies and Sa spirals), as well
as very late-type spirals (Sd and beyond).  Some galaxies near these
endpoints are included in the Frei sample, but some bar classes are
missing: strongly barred SBab galaxies are absent from the sample,
while all Scd galaxies in the sample are weakly barred SABcd systems.

Fortunately, the Frei et al.\ (1996) sample is most representative
(both in terms of relative numbers of systems as a function of
Hubble stage, and in terms of the mix of SA, SAB, and SB systems)
for the intermediate--late type spirals (types Sb, Sbc, and Sc)
that define the two arms of the original tuning fork.  While the
absolute number ($\sim 50$) of such systems is not large, it is
certainly comparable to the total number of archetypal galaxies
that define the visual classification sequence.  Thus, the Frei
et al.\ (1996) sample, while modest in size, is actually a rather
useful probe of the bifurcation of intermediate--late spiral
galaxies into barred and unbarred systems, though it must be used
with caution when probing early-type and very late-type spirals.

Figure~\ref{fig:examples} illustrates the inner and outer ellipses
determined for three representative barred spirals in this sample
using the multi-thresholding technique described above.  These objects
are illustrative of the general success of the multi-thresholding
technique in isolating galactic bars over broad range of intensity and
size. The orientation and semi-major axis length of the bar is
determined very robustly by this technique. On the whole, bar axial
ratios are also well-constrained, although they tend to be somewhat
underestimated for early type systems, because the contribution from a
strong central bulge tends to fatten inner isophotes and make bars
less well-defined.  In fact, a similar effect also impacts visual
classifications, as is clearly seen in Figure~\ref{fig:rc3frei} --
note the sharp increase in the proportion of systems in the RC3 with
unknown bar classifications as one moves from late-type to early-type
galaxies.

The distribution of this sample in Hubble space is shown in
Figure~\ref{fig:stamps}.  Small ``postage-stamp'' images of the
individual galaxies denote the position of each object in
parameter space, while the surrounding colored boxes divide the
galaxy population into SA (red), SAB (green), and SB (blue)
spirals. For comparison, this figure also shows the Hubble space
positions of the elliptical galaxies in the Frei et al.\ (1996)
sample, surrounded by black boxes.

Several points are evident from inspection of Figure~\ref{fig:stamps}:
\begin{enumerate}
\item Barred, weakly-barred, and unbarred galaxies are remarkably
well-separated in the diagram.  The quantitative parameters
defining Hubble space are clearly rather closely related to the
subjective criteria upon which the original tuning fork
classification system was based. There is no intermixing of
unbarred galaxies and strongly barred galaxies, and only a small
amount of intermixing between strongly barred and weakly barred
galaxies. A close inspection of the individual galaxy images
shown in Figure~\ref{fig:stamps} suggests that even this small
amount of mixing may well be due to visual misclassification. The
single elliptical galaxy scattered into the SB distribution is the
result of an image defect (a slight ramp in the sky background).
\item There is a general broad correlation between $C$ and \fb.  This
trend presumably just reflects the diluting effects of large bulges on
quantitative measures of bar strength, discussed above.
\item The spiral sequence is clearly bimodal.  This bimodality is
better displayed in Fig.~\ref{fig:hist}, which shows the
distribution of spiral galaxies in Fig.~\ref{fig:stamps}
projected on to an axis perpendicular to the mean correlation
between $C$ and \fb.  Strongly barred spirals do not simply
constitute the boundaries of a smooth unimodal distribution of
bar strength.  Instead, barred and unbarred spirals appear to
populate two parallel sequences, with remarkably few galaxies
occupying the region in between. It would be useful to assess the formal significance of the bimodality
apparent in Fig.~4, using, for example, the KMM test (Ashman, Bird \& Zepf
1994).  However, there are two reasons why such a test would not yet be
appropriate.  First, this distribution was derived by projecting Fig.~3
along the axis which maximizes the bimodality signal, and this
maximization would not be accounted for in the significance of the
test.  Second, the current data set has folded in with it the selection
that Frei et al.\ made in defining their sample, and we have no way of
quantitatively assessing the impact of this selection on the shape of this
distribution.  Nonetheless, it will be important to carry out such a test
when a larger, more objectively-defined sample becomes available.
\item Rather surprisingly, weakly barred systems do not, on the whole,
seem to correspond to systems bridging the parameter space between
unbarred and strongly barred galaxies. Weakly barred SAB galaxies
seem to be an extension of the SA sequence.
\end{enumerate}

\begin{figure*}[bht]
  \centering  \epsfig{figure=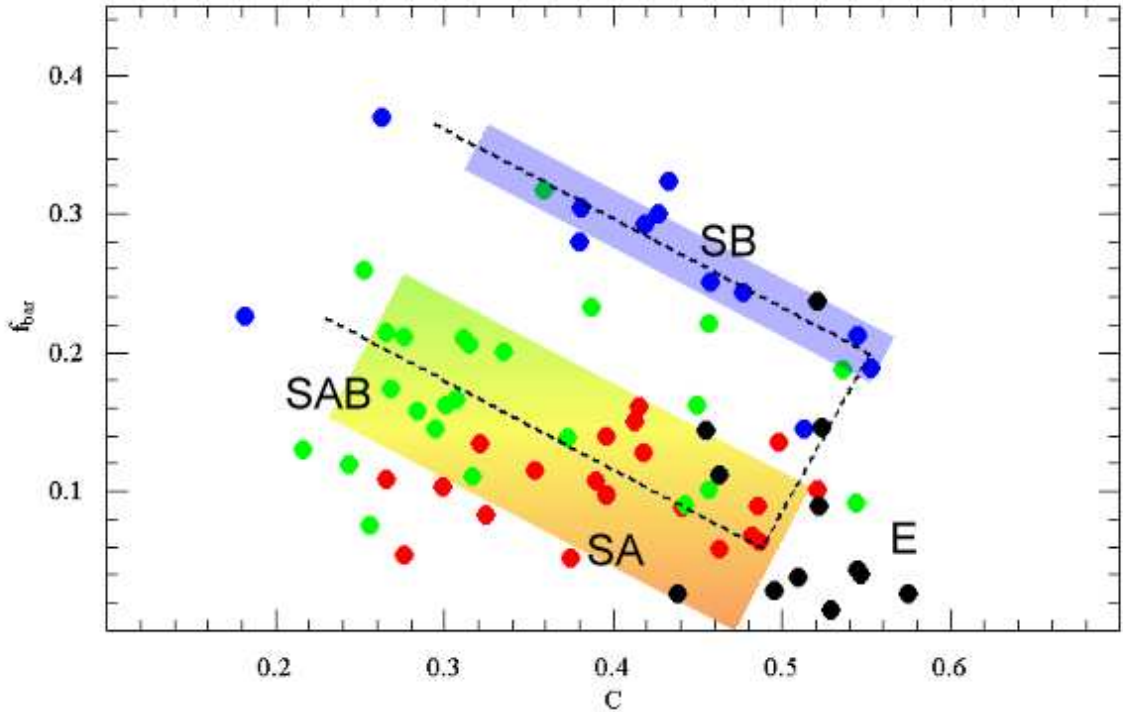,width=6.5in}
  \caption{ The distribution of galaxies shown in
  Figure~\ref{fig:stamps}. Unbarred galaxies are shown in red,
  weakly barred galaxies are shown in green, strongly barred
  systems are shown in blue, and elliptical galaxies are shown in
  black. The dashed line indicates a schematic tuning fork.
  Shaded regions in the figure correspond to 1$\sigma$
  rectangular contours delineating the two tines of the
  quantitative tuning fork. }  \label{fig:regions}
\end{figure*}

\section{DISCUSSION}\label{sec:discussion}

A schematic overview of Hubble space is presented in
Fig.~\ref{fig:regions}, which shows where various classes of galaxies
lie in this morphological parameterization.  Colored rectangles on
this diagram correspond to approximate ``1$\sigma$'' contours in
parameter space for the various populations.  The dashed line
intersecting these regions gives a schematic representation of the
classical tuning fork within this parameter space.  What is most
remarkable about this diagram is how apt Hubble's tuning fork
description remains almost seventy-five years after its introduction.
The correspondence is not perfect: in Hubble space, the tuning fork is
both skewed and horizontally flipped relative to its usual
representation, and elliptical galaxies are not arrayed in a sequence
of axial ratio.  However, the bimodal distribution of galaxies is
clearly rather more than a convenient textbook abstraction of a smooth
continuum in the bar strength of galaxies. There really is a genuine
bifurcation in the properties of galaxies, which cleanly divides
barred and weakly/unbarred galaxies.  The fact that SB galaxies lie on
a remarkably sharply-defined sequence strongly suggests the presence
of some underlying unifying physical process -- perhaps the classical
bar instability (Binney \& Tremaine 1987, and references therein).

The Frei et al.\ (1995) sample indicates that weakly-barred SAB
galaxies appear much more closely related to SA galaxies than to SB
galaxies, with the SABs forming the late-type extension of the SA
branch of the tuning fork.  It is important to avoid over-interpreting
this result, since it partially depends on the mix of later-type
galaxies for which, as we have discussed, the Frei sample is not
entirely representative. Nonetheless, it seems quite possible that
this will prove to be an example of a phenomenon that is obvious in a
quantitative analysis, but which is difficult to detect using
subjective criteria.  Furthermore, it may well explain why the
proportion of weakly barred systems is very discrepant in local
catalogs.  For example, there is reasonable agreement in local
catalogs that the proportion of strongly barred galaxies is
25\%--35\%, based on the numbers given in the RC3, the {\em Revised
Shapley-Ames Catalogue} (RSA; Sandage \& Tammann 1987),and the {\em
Uppsala General Catalogue} (UGC; Nilson 1973).  However, an additional
30\% of spirals are classed as weakly barred in the RC3, substantially
higher than in the UGC or RSA.  Figure~\ref{fig:regions} suggests that
this discrepancy is manifestation of the non-fundamental nature of the
``bars'' seen in these objects, rendering them exceptionally
problematic to pigeonhole visually.

Further support for the link between SAs and SABs can, in retrospect,
be seen directly from the RC3 data.  As Fig.~\ref{fig:rc3frei} shows,
the fraction of SAB galaxies grows and the fraction of SA galaxies
decreases as one goes along the Hubble sequence from S0 to Sd.  Again,
it is tempting to invoke a simple underlying physical process.
Perhaps, for example, the small bulges and weak bars of the SAB
galaxies are both the result of the bar buckling instability seen in a
number of numerical simulations (Combes \& Sanders 1981, Raha et al.\
1991).  Such a process could well leave a modest non-axisymmetric
signature in what is essentially an unbarred galaxy, explaining the
close relation of these systems to the SAs (Kormendy 1992).

Variations in the distribution of galaxies in Hubble space as a
function of rest wavelength will be investigated in a future
paper, but it seems likely that the distribution will be at least
moderately wavelength-dependent. Understanding the nature of this
wavelength dependence will be important in order to extend our
techniques for use in studying galaxies at higher redshifts. For
example, Eskridge et al. (2000) found a higher fraction
of barred galaxies when observing at longer wavelengths, which
could produce a spurious decrease in the fraction of barred
galaxies with redshift if one were to observe a sample of
intrinsically identical galaxies through a single filter.
Extending the methodology introduced in this paper to samples
imaged at a range of wavelengths will allow direct comparison
between the properties of nearby galaxies and those in
higher-redshift samples.

\section{CONCLUSIONS}\label{sec:conclusions}

Recent studies of high-redshift galaxies have revitalized the
entire field of galaxy morphology.  The myriad new forms of
galaxies visible on deep HST images has led morphologists to
devise new quantitative galaxy classification systems that,
insofar as they connect to the classical Hubble system at all,
focus on the early-to-late classification sequence and neglect
the distinction between barred and unbarred galaxies.  However,
dynamical studies and direct observations of variations in the
fraction of barred galaxies with redshift clearly imply that the
bifurcation into barred and unbarred systems is a fundamental
aspect of galaxy evolution.

In this paper, our goal has been to present a framework for combining
these factors.  We define a quantitative two-dimensional morphological
parameterization (``Hubble space'') that maps closely on to Hubble's
original subjective visual definition of the morphological tuning
fork.  Using the Frei et al.\ (1995) sample, we have made a
preliminary foray into Hubble space.  We find that the distribution of
bar strengths in galaxies in this sample is sharply bimodal --- barred
spirals do not simply delineate the extrema in a nearly uniform
progression of galactic bar strength.  Overall, this work lends
support to Hubble's original picture in which a genuine physical gulf
exists between the properties of barred and regular galaxies, and
indicates that a natural division in parameter space can be used to
distinguish barred from regular spirals objectively.

In addition, this work also shows how adopting a quantitative approach
sheds light on a possible close link between regular and weakly-barred
spirals that is obscured by the nomenclature of visual classification.
This analysis indicates that weakly barred systems may be naturally
viewed as an extension of the regular spiral sequence, rather than as
bridges between strongly barred and regular spiral systems.  This
conclusion is based on a small sample of galaxies, however, and the
relationship between various categories of barred spiral galaxies
clearly needs to be investigated in a more thorough manner with a much
larger, more suitably-selected sample.  In any case, the rich variety
of galaxy forms visible on deep HST images shows that the parameters
defining Hubble space are insufficient, on their own, to fully
encompass galaxy morphologies at high redshifts.  Nevertheless, this
investigation demonstrates that parametric measures of bar strength
should certainly be incorporated into whatever successor to Hubble's
sequence is ultimately adopted in order to describe galaxy
morphologies over a broad range of look-back times.

Hubble constructed his original tuning fork guided strongly by his
intuition and experience, but knowing comparatively little about
the dynamical construction of galaxies. Seventy-five years later,
Hubble's intuition stands vindicated, with the broad outline of
his tuning fork holding up remarkably well to quantitative
inspection.

\acknowledgments We thank Sidney van den Bergh, Richard Ellis, and
Jarle Brinchmann for many valuable discussions on the physical origins
of galaxy morphology. This research was supported by a grant from
NSERC.


\begin{thebibliography}{1}

\bibitem[]{abr94}
Abraham, R.~G., Valdes, F., Yee, H.~K.~C. \& van den Bergh, S. 1994,
ApJ, 432, 75

\bibitem[]{abr96a}
Abraham, R.~G., Tanvir, N.~R., Santiago, B.~X., Ellis, R.~S., Glazebrook, K.,
  van~den Bergh, S. 1996a, MNRAS, 279, L47

\bibitem[]{abr96b}
Abraham, R.~G., van~den Bergh, S., Ellis, R.~S., Glazebrook, K.,
Santiago, B. X., Griffiths, R.~E., Surma, P. 1996b, ApJS, 107, 1

\bibitem[]{abr99} Abraham, R. G., Merrifield, M. R., Ellis, R. S., Tanvir,
  N. R., \& Brinchmann, J. 1999. MNRAS, 308, 569.

\bibitem[]{ash94} Ashman, K., Bird, C., \& Zepf, S. 1994. AJ, 108, 2348.

\bibitem[]{ber00} Bershady, M.~A., Jangren, A. \& Conselice, C.~J. 2000, AJ,
119, 2645

\bibitem[]{bin87}
Binney, J. \& Tremaine, S., 1987, Galactic Dynamics (Princeton: Princeton
University Press)

\bibitem[]{bri98} Brinchmann, J., Abraham, R. G., Schade, D., Tresse, L.,
Ellis, R. S., Lilly, S. J., Le Fevre, O., Glazebrook, K.,
Hammer, F., Colless, M., Crampton, D., \& Broadhurst, T.
1998, ApJ, 500, 75.

\bibitem[]{cas90} Casertano, S. \& van Albada, T.S., 1990, in Bayonic Dark
Matter, eds D. Lynden-Bell \& G. Gilmore (Dordrecht: Kluwer), 159

\bibitem[]{com81} Combes, F. \& Sanders, R.H., 1981, A\&A, 96, 164


\bibitem[]{cor2000} {Corbin}, M.\ R., {Vacca}, W.\ D., {O'Neil}, E.,
{Thompson}, R.\ I., {Rieke}, M.\ J. \& {Schneider}, G. 2000. AJ, 119, 1062

\bibitem[]{dev91} de~Vaucouleurs, G., de~Vaucouleurs, A., Corwin, H.G., Buta, R.J.,
Paturel, G. \& Fouqu/'e, P., 1991, Third Reference Catalogue of Bright
Galaxies (New York: Springer-Verlag)

\bibitem[]{doi93} Doi, M., Fukugita, M., \& Okamura, S. 1993, MNRAS, 264, 832.

\bibitem[]{dri95} {Driver}, S.~P., {Windhorst}, R.~A., {Griffiths}, R.~E. 1995a, ApJ, 453, 48

\bibitem[]{dri98} Driver, S. P., Fernandez-Soto, A., Couch, W. J.,
Odewahn, S. C., Windhorst, R. A., Lanzetta, K., \&
Yahil, K. 1998, ApJ(Lett), 496, 93


\bibitem[Eskridge et al.\ (2000)]{2000AJ....119..536E} Eskridge, P.\ B.\
and 11 colleagues 2000, \aj, 119, 536

\bibitem[]{fre96} Frei, Z., Guhathakurta, P., Gunn, J.E., Tyson, J.A., 1996,
AJ, 111, 174

\bibitem[]{gla95}{Glazebrook}, K., {Ellis}, R., {Santiago}, B., {Griffiths},
R. 1995, MNRAS,  275, L19

\bibitem[]{gou97} Gould, S. J., 1998. ``The Clam Stripped Bare by Her
Naturalists, Even'', in {\em Leonardo's Mountain of Clams and the Diet
of Worms}, (Three Rivers Press: NY), p. 96.

\bibitem[]{gri94} {Griffiths}, R.\ E., {Casertano}, S., {Ratnatunga}, K.\ U.m
    {Neuschaefer}, L.\ W., {Ellis}, R.\ S., {Gilmore}, G.\ F.,
    {Glazebrook}, K., {Santiago}, B., {Huchra}, J.\ P.,
    {Windhorst}, R.\ A., {Pascarelle}, S.\ M., {Green}, R.\ F.,
    {Illingworth}, G.\ D., {Koo}, D.\ C. \& {Tyson}, A.\ J. 1994,
    ApJ(Lett), 435, L19

\bibitem[]{has90} Hasan, H. \& Norman, C., 1990, ApJ, 361, 69


\bibitem[]{hub26} Hubble, E. 1926. ApJ, 64, 321

\bibitem[]{kor92} Kormendy, J. 1992, in Proc.IAU Symp.153, ``Galactic Bulges'',
p.209, Kluwer, Dordrecht, eds. Dejonghe, H., Habing, H.

\bibitem[]{mar98} Marleau, F. R. \& Simard, L. 1998, ApJ, 507, 585

161, 903

\bibitem[]{nai95} Naim, A. Lahav, O., Buta, R.J., Corwin, H.G., de Vaucouleurs, G.,
Dressler, A., Huchra, J.P., van den Bergh, S., Raychaudhury, S., Sodre, L.,
Storrie-Lombardi, M.C., 1995, MNRAS, 274, 1107

\bibitem[]{} Nilson, P. 1973. {\em Uppsala General Catalogue of
Galaxies}, Acta Upsaliensis Ser V: A Vol I


\bibitem[]{ode96}
{Odewahn}, S.~C., {Windhorst}, R.~A., {Driver}, S.~P., {Keel}, W.~C. 1996,
ApJL, 472, L13

\bibitem[]{pfe91} Pfenniger, D., 1991, in Dynamics of Disk Galaxies, ed.\
B. Sundelius (G\"oteborg: G\"oteborgs University), 191

\bibitem[]{phi97} {Phillips}, A.\ C., {Guzman}, R., {Gallego}, J., {Koo}, D.\ C.,
    {Lowenthal}, J.\ D., {Vogt}, N.\ P., {Faber}, S.\ M. \&
    {Illingworth}, G.\ D. 1997. ApJ, 489, 543.

\bibitem[]{rah91} Raha, N., Sellwood, J.A., James, R.A., Kahn, F.D., 1991,
Nature, 352, 411

\bibitem[]{san87} Sandage, A. \& Tammann, G.A., 1987, A Revised Shapley-Ames
Catalog of Bright Galaxies (Carnegie Institute of Washington)

\bibitem[]{sch95} {{Schade}, D., {Lilly}, S.\ J., {Crampton}, D., {Hammer}, F.,
    {Le F\'evre}, O. \&{Tresse}, L.}, ApJ(Lett), 451, 1

(astro-ph/9807010)

\bibitem[]{vdb96} van den Bergh, S., Abraham, R. G.,
Ellis, R. S., Tanvir, N. R., Santiago, B. X. 1996, AJ, 112, 359.

\bibitem[]{vdb00} van den Bergh, S., Cohen, J.~G., Hogg, D.~W. \&
Blandford, R. 2000, AJ, in press (astro-ph/0008051)

\bibitem[]{vol2000} {Volonteri}, M., {Saracco}, P. \& {Chincarini}, G. 2000.
A\&A(Supp), 145, 111.

\bibitem[]{wil96} Williams et al.\ 1996, AJ, 112, 1335

\end{thebibliography}
\end{document}